\documentclass[fleqn,twoside]{article}
\usepackage{espcrc2}
\usepackage{amsmath,amssymb}

\readRCS
$Id: espcrc2.tex,v 1.2 2004/02/24 11:22:11 spepping Exp $
\usepackage{graphicx}
\usepackage[figuresright]{rotating}

\def\Bbar{\overline{B}^0}
\def\Abar{\overline{A}}
\def\Sbar{\overline{S}}
\def\lambdaB{\bar{\lambda}}
\def\GammaB{\overline{\Gamma}}

\newcommand{\beq}{\begin{equation}}
\newcommand{\eeq}{\end{equation}}
\newcommand{\beqa}{\begin{Eqnarray}}
\def\eeqa{\end{Eqnarray}}

\title{Obtaining $\alpha$ from $B\to \rho^\pm \pi^\mp$\thanks{Talk presented at the Sixth International
 Conference on Hyperons, Charm and Beauty Hadrons, IIT, Chicago, June 27 - July 3, 2004.}}

\author{J. Zupan \address[Technion]{Department of Physics,
Technion--Israel Institute of Technology,
Technion City, 32000 Haifa, Israel}${}^{,}$\address{J.~Stefan Institute, Jamova 39, P.O. Box 3000,1001
Ljubljana, Slovenia}
        }
       
\begin{document}

\begin{abstract}
We discuss how to extract the weak phase $\alpha$ from present
data on   $B\to \rho^\pm \pi^\mp$ decays. Introducing $\alpha_{\rm eff}$ and
constraining the difference from $\alpha$ using flavor SU(3), one
arrives at $\alpha=(95\pm 16)^\circ$, if a testable assumption of a small
 relative strong phase between the two relevant tree amplitudes is employed to distinguish between discrete ambiguities. On
the long run we advocate the combined fit to $B\to \rho^\pm \pi^\mp$
and SU(3) related modes. The effect of SU(3) breaking in this
approach is expected to be small, because of relatively small penguin
pollution.
\vspace{1pc}
\end{abstract}

\maketitle

\section{Introduction}
The difficulty in extracting CKM angle $\alpha$ from time dependent
measurements of $b\to u \bar{u} d$ decays is that the penguin
contributions to the decays are in general nonnegligible and have to
be taken into account. There are two basic approaches to this problem
in the case of $\rho^\pm \pi^\mp$ final state,
both dating more than a decade ago. The isospin analysis
approach~\cite{Lipkin:1991st,Gronau:1991dq} is an extension of the
original isospin triangle relations approach to extracting $\alpha$ in
$B\to\pi\pi$ system~\cite{Gronau:1990ka}.
 A complication with the $\rho \pi$ final state is that the 
 analysis requires a construction of 
two pentagons.  Since none of the sides of pentagons seem to be much smaller than the
others, thus simplifying the analysis,  a useful measurement of
$\alpha$ using the full isospin analysis may be impractical even with super-B-factory-like luminosities \cite{Stark:2003nq}.

A  more promising way of learning $\alpha$ in these 
decays is based on performing a time-dependent Dalitz plot analysis of  $B^0 \to 
\pi^+\pi^-\pi^0$~\cite{Snyder:1993mx}, which allows one to obtain the phase differences of decay
amplitudes from  the interference between 
two $\rho$ resonance bands. This raises issues such as
the precise shapes of the tails of the Breit-Wigner functions, and
the effect of interference with other resonant and non-resonant
contributions~\cite{Deandrea:2000tf}. A complete implementation of this method
requires higher statistics than is available today. 

In this
talk we will try to answer a more modest question, namely what one can
learn about  $\alpha$ from the data available at this
moment? The first obstacle that one encounters in extracting $\alpha$ from the  time-dependent decay measurements of  
$B^0(\Bbar)\to \rho^\pm \pi^\mp$ by the BABAR~\cite{BaBar} and BELLE~\cite{Belle,Wang:2004va} 
collaborations is that
these processes involve more hadronic parameters than measurable 
quantities. Further assumptions are therefore required to answer the question in a
model-independent manner. We will use flavor SU(3), a symmetry less precise 
than isospin, to  relate 
$B^0\to \rho^\pm\pi^\mp$ to processes of the type $B\to K^*\pi$ and $B \to 
\rho K$ \cite{Gronau:2004tm,HLLW,CKMfitter}. 

\section{Observables}
We start by setting up notations and conventions. The decay
amplitudes are denoted by
\beq
\begin{split}
A_\pm=A(B^0\to \rho^\pm\pi^\mp)=& {e^{i \gamma}t_{\pm}}+{p}_{\pm},~\\
\overline{A}_\pm =A(\Bbar\to \rho^\mp\pi^\pm)=&e^{- i\gamma} t_{\pm}+
p_{\pm}.
\label{amplitudes}
\end{split}
\eeq
Each of the four amplitudes can be decomposed in two terms,
a``tree" ($t_\pm$) and a 
``penguin" ($p_\pm$) amplitude, carrying specific CKM factors. The
tree amplitude is proportional to $V^*_{ub}V_{ud}$ by definition,
giving dependence on the weak phase $\gamma$ in Eq.~\eqref{amplitudes}.  Note that the amplitudes 
$t_+ (p_+)$ and $t_-(p_-)$ have different dynamical origins and are expected
to involve different magnitudes and different strong phases.  We define three strong phase 
differences
\beq\label{delta}
\delta_\pm=\arg\big(p_\pm/t_\pm\big)~,\qquad \delta_t=\arg\big(t_-/t_+\big)~.
\eeq
In addition, we also define the ratios 
\beq
r_\pm \equiv \left|{p_\pm}/{t_\pm}\right|~,~~~~~~r_t \equiv \left|{t_-}/{t_+}\right|~.
\eeq
Counting parameters, we find a total of 8, consisting of 7 hadronic quantities 
$|t_{\pm}|,|p_{\pm}|, \delta_{\pm},\delta_t$ and the weak phase
$\gamma$ or $\alpha$ ($\beta$ is assumed to be known). 

\begin{figure}
\begin{center}
\includegraphics[width=3.5cm]{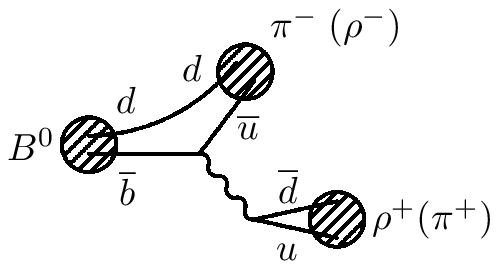}
\includegraphics[width=3.5cm]{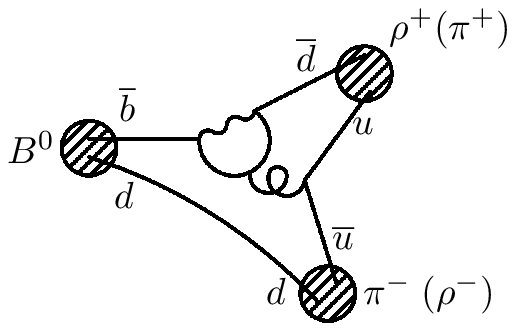}\\[-7mm]
\caption{\footnotesize{The tree (left) and penguin (right) diagrams
for the $B^0\to \rho^+\pi^-$ ($B^0\to \rho^-\pi^+$) decays.\vspace{-9mm}}} \label{figDecay}
\end{center}
\end{figure}

Let us now consider measurables in time-dependent rates.  
Time-dependent decay rates for initially $B^0$ decaying into 
$\rho^\pm\pi^\mp$ are given by
\begin{equation*}
\begin{split}
&\Gamma(B^0(t) \to \rho^\pm\pi^\mp) = e^{-\Gamma t} \frac{1}{2}\Gamma^{\rho\pi} \left (1 \pm {\cal A}_{\rm CP}^{\rho\pi}\right )\times\\
&\times\left [ 1 + (C \pm \Delta C)\cos\Delta mt 
- (S \pm \Delta S)\sin\Delta mt\right ].
\end{split}
\end{equation*}
 For initially $\Bbar$ decays, the  $\cos\Delta mt$ and $\sin\Delta mt$ terms
 reverse signs. There are 6 measurables:
$C$, $\Delta C$, $S$, $\Delta S$,  $\Gamma^{\rho\pi}$, ${\cal A}_{\rm
CP}^{\rho \pi}$ that are  parametrized by  8 unknowns discussed
above. Thus it is not possible to extract $\alpha$ without
further assumptions, which will be made explicit shortly. A
compilation of experimental data can be found in \cite{Gronau:2004tm}.


\subsection{More economical set}
We now introduce a smaller set of observables that still contains all
the information needed to extract $\alpha$. Expanding in terms of
penguin-to-tree ratios, $r_\pm$, one arrives at
\begin{align}
S=&\frac{2 r_t}{1+r_t^2}\sin 2\alpha \cos\delta_t+O(r_\pm),\\
\Delta S= & \frac{2 r_t}{1+r_t^2}\cos 2\alpha \sin\delta_t +O(r_\pm).
\end{align}
One can get rid of $|t_\pm|$ (and $r_t$),
if one considers only the ratios of decay widths and appropriately 
rescaled $S$, $\Delta S$ such that no dependence on $r_t$ occurs. This
leads us to the smaller set of 2 direct CP asymmetries
$$
{\cal A}^{\pm}_{\rm CP} \equiv \left(|\Abar_\pm|^2 - |A_\pm|^2\right)/\left(|\Abar_\pm|^2 + |A_\pm|^2\right),
$$
and 2  observables related to $S$ and $\Delta S$  
\begin{equation*}
\begin{split}
\begin{Bmatrix}
\Sbar\\ 
\Delta \Sbar
\end{Bmatrix}
\propto\Big[&\big(S+\Delta
S\big)\left(\frac{1+{\cal A}_{CP}^{\rho\pi}}{1-{\cal A}_{CP}^{\rho\pi}}\right)^{1/2}\pm
\\
&
\big(S-\Delta
S\big)\left(\frac{1-{\cal A}_{CP}^{\rho\pi}}{1+{\cal A}_{CP}^{\rho\pi}}\right)^{1/2}
\Big],
\end{split}
\end{equation*}
where the undisplayed common normalization factor can be found in
\cite{Gronau:2004tm}. Expanding in  $r_\pm$ the leading terms are
\begin{align}
\Sbar=&\sin 2\alpha \cos\delta_t+O(r_\pm)\label{Sbar},\\
\Delta \Sbar= & \cos 2\alpha \sin\delta_t + O(r_\pm)\label{DeltaSbar},\\
{\cal A}^{\pm}_{\rm CP} =& -2r_{\pm}\sin\delta_{\pm}\sin(\gamma)+ O(r_\pm^2)\label{ApmCP}.
\end{align}
The above set of 4 observables  depends on 6 unknowns: $\delta_\pm$, $r_\pm$, $\delta_t$, and
$\alpha$. To measure $\alpha$ one has to know $O(r_\pm)$ terms in $\Sbar$, $\Delta
\Sbar$. There are three basic approaches to this problem, one can
either (i) bound corrections, which we will do by constructing
$\alpha_{\rm eff}$, (ii) measure corrections, which will be done
through SU(3) related modes, or (iii) calculate corrections, for
instance in the framework of  QCD factorization \cite{BN}.
\section{SU(3) related modes}
When relating the tree amplitudes we also account for the SU(3)
breaking using guidance from factorization
\begin{align}
t'_+  &=  \frac{f_{K^*}}{f_\rho}\frac{V^*_{ub}V_{us}}{V^*_{ub}V_{ud}}\,t_+ = 
\frac{f_{K^*}}{f_\rho}\lambdaB\,t_+,\\
t'_- & =  \frac{f_K}{f_\pi}\frac{V^*_{ub}V_{us}}{V^*_{ub}V_{ud}}\,t_- = 
\frac{f_K}{f_\pi}\lambdaB\,t_-,
\end{align}
while for the penguin amplitudes we assume exact SU(3) relations for
the moment
\beq
p'_{\pm}  =  \frac{V^*_{cb}V_{cs}}{V^*_{cb}V_{cd}}\,p_{\pm}
= -\lambdaB^{-1}p_{\pm}.
\eeq
Note that in deriving the above relations annihilation like
topologies were neglected. Since the tree amplitudes in the $\Delta
S=1$ modes are $\lambdaB$ suppressed, while the penguins are enhanced,
they are perfect probes for the sizes of penguin contributions. Of
interest are the following ratios of the CP averaged decay widths:
ratios with the
``tree +penguin'' $\Delta S=1$ decays
${\cal R}^0_+  \equiv  {\lambdaB^2\GammaB(B^0\to K^{*+}\pi^-)}/
{\GammaB(B^0 \to \rho^+\pi^-)}$, $
{\cal R}^0_-   \equiv {\lambdaB^2\GammaB(B^0\to \rho^- K^+)}/
{\GammaB(B^0\to \rho^-\pi^+)}$, and ratios with the ``only penguin''
$\Delta S=1$ decays
 $
{\cal R}^+_+  \equiv  {\lambdaB^2\GammaB(B^+\to K^{*0}\pi^+)}/
{\GammaB(B^0 \to \rho^+\pi^-)}$,
 ${\cal R}^+_-  \equiv  {\lambdaB^2\GammaB(B^+\to \rho^+ K^0)}
{\GammaB(B^0\to \rho^-\pi^+)}$, of which only
 ${\cal R}^+_-$ has not been measured yet.

\subsection{Bounds on penguins}
These ratios can be used to bound  $r_\pm$. For $r_+$ the strictest
bound at present is obtained from
the ``penguin only'' ratio ${\cal R}^+_{+}$ that constrains $r_+$ to
be in the range 
$$
{\sqrt{{\cal R}^+_{+}}}/\left({1 + \sqrt{{\cal R}^+_{+}}}\right)
\le  r_{+}  \le 
{\sqrt{{\cal R}^+_{+}}}/{\left( 1 - \sqrt{{\cal R}^+_{+}}\right)},
$$
which gives \cite{Gronau:2004tm}
\beq
0.14~(0.16) \le r_+ \le 0.24~(0.21),
\eeq
at 90 \% CL (values in parantheses are obtained by using central
values of ${\cal R}^+_{+}$). Symilarly one obtains for $r_-$ \cite{Gronau:2004tm}
\beq
0.12(0.21) \le r_- \le 0.32~(0.27).
\eeq
Note that slightly lower values were predicted in
QCD factorization \cite{BN}
\beq
r_+ = 0.10^{+0.06}_{-0.04} {\rm ~~~and~~~} r_- = 0.10^{+0.09}_{-0.05}.
\eeq

\section{$\alpha_{\rm eff}$ and $\alpha$}
Let us now turn to the initial question of bounding $\alpha$ from
presently available data. Following $B(t)\to \pi^+\pi^-$ (see
e.g. \cite{Gronau:2004gd}) we define
\beq
\alpha^{\pm}_{\rm eff} \equiv \frac{1}{2}\arg\left (e^{-2i\beta}\Abar_{\pm}A^*_{\pm}\right ),
\eeq
which are {\it not} observables. The observables on the contrary mix
$\Abar_+$ and $A_-$ amplitudes and are
\beq\label{alphaRel}
\begin{split}
2\alpha^{\pm}_{\rm eff} \pm \hat\delta &\equiv \arg\left
(e^{-2i\beta}\Abar_{\pm}A^*_{\mp}\right ) \\
&= 
\arcsin\left (\frac{S\pm \Delta S}{\sqrt{1- (C\pm \Delta C)^2}}\right),
\end{split}
\eeq
with $
\hat\delta =\delta_t + O(r_\pm)
$. The average of the above observables we define to be $\alpha_{\rm eff}$
\beq\label{alphaeff}
 \alpha_{\rm eff} \equiv \frac{1}{2}\left (\alpha^{+}_{\rm eff} + \alpha^{-}_{\rm eff} \right )= \alpha +O(r_\pm) .
\eeq
To have a handle on $\alpha$ we therefore have to estimate $O(r_\pm)$
corrections in \eqref{alphaeff} which are bounded by SU(3) related
modes.  The strongest bound in the case of $\alpha_{\rm eff}^+$ comes
from ``penguin only'' ratio, for which at 90\% CL
\begin{equation*}
\begin{split}
|\alpha^{+}_{\rm eff} - \alpha| &=\frac{1}{2}\arccos\left(\frac{1 - 2{\cal R}^+_{+}\sin^2(\beta + \alpha)}
{\sqrt {1 - {\cal A}^{+ 2}_{CP} } }\right)  \\
&\le 7.1^\circ - 11.5^\circ.
\end{split}
\end{equation*}
The equivalent of Charles bound~\cite{Charles} also exists 
\begin{equation*}
|\alpha^{-}_{\rm eff} - \alpha | \le \frac{1}{2} \arccos\left(
\frac{1 - 2{\cal R}^0_{-}}{\sqrt {1 - {\cal A}^{- 2}_{\rm CP}}}\right) \le 14.9^\circ.
\end{equation*}
Putting the bounds together we have
\beq\label{Delta-alpha}
|\alpha_{\rm eff} - \alpha| \le 11.0^\circ - 13.2^\circ.
\eeq
This implies that we can get a value for $\alpha$ now!
Using a mild assumption, testable from partial Dalitz
plot analysis \cite{Gronau:2004tm},  that $|\delta_t|\ll 90^\circ$ ~~(QCD
factorization gives $\delta_t=(1\pm3)^\circ$ \cite{BN}) one can distinguish  ambiguous
solutions for $\alpha^\pm_{\rm eff}\pm\hat\delta$. Namely the difference
\beq
(2\alpha^+_{\rm eff} +\hat\delta) - (2\alpha^-_{\rm eff} - \hat\delta) = 2\delta_t + 
{\cal O}(r_{\pm}),
\eeq
has to be small ($\ll 180^\circ$). This leaves the following viable solutions
in the $(0^\circ,180^\circ)$ range
\beq
\alpha_{\rm eff} =  \{95^\circ\pm 6^\circ, 175^\circ\pm 6^\circ\}
\Rightarrow \alpha=
(95 \pm 16)^\circ 
\eeq
with the last error incorporating also an estimate of
SU(3) breaking effects. Note also, that the extracted value of
$\alpha$ is most  sensitive to changes in $S$, and not very sensitive
to $\Delta S$.

\begin{figure}
\begin{center}
\includegraphics[width=6.5cm]{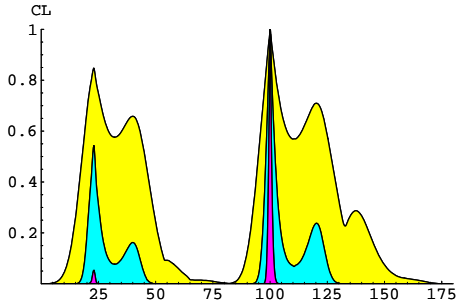}\\[-9mm]
\caption{\footnotesize{Confidence level (CL) as a function of $\alpha$ for a generated
set of data (for details see \cite{Gronau:2004tm}). 
Errors used for
$\chi^2$ are the currently measured ones [yellow (light gray) region], those anticipated with 
ten times statistics [cyan (gray)] and hundred times statistics [purple (dark gray)].\vspace{-9mm} }} \label{FigPValue}
\end{center}
\end{figure}

\section{SU(3) fit}
Finally we discuss how one can obtain $\alpha$ model independently
(thus relaxing the mild assumption of $|\delta_t|\ll 90^\circ$,
but still neglecting annihilation). Adding SU(3) related modes we have
in addition to observables $\Sbar$, $\Delta \Sbar$,
 ${\cal A}^{\pm}_{\rm CP} $ also the ratios
\begin{align}
\hspace{-0.6cm}\frac{1}{2}\left({\cal R}_\pm^0+{\cal R}_\pm^+\right)=&r_\pm^2+\dots\label{Rone}\\
\hspace{-0.6cm}\frac{1}{2}\left({\cal R}_\pm^+-{\cal R}_\pm^0\right)=& r_\pm
\bar{\lambda}_\pm^2 \cos(\delta_\pm)\cos(\gamma)+\dots \label{Rtwo}
\end{align}
giving a total of 8 measurables, and only  6 unknowns: $\delta_\pm$,
$r_\pm$, $\delta_t$, $\alpha$, and therefore consitute an
overconstrained system of equations. In Fig. \ref{FigPValue} we show
an example where data are generated with input parameters
$\delta_t=170^\circ$, $\alpha_{\rm input}=100^\circ$ and other
parameters in experimentally  permited ranges (for details see
\cite{Gronau:2004tm}). One sees that (i) there exist
ambiguities in the solutions of Eqs. \eqref{Sbar}-\eqref{ApmCP},
\eqref{Rone}, \eqref{Rtwo}, but (ii) eventually with enough statistics only one solution remains.

The (dis-)apperance of ambiguities can be followed step by step in the
$r_\pm$ expansion. At leading order there is a 16-fold ambiguity
(cf. Eqs. \eqref{Sbar}, \eqref{DeltaSbar}). This is resolved by
higher order terms in $r_\pm$,  meaning that the 
observables have to be measured with $O(r_\pm)\sim 20\%$ precision or
better.

A positive aspect of relatively small $r_\pm$, on the other hand, is
that the effect of SU(3) breaking on extracted $\alpha$ will be small,
effectivelly of order
$r_\pm^2$.  A Monte Carlo study incorporating  up to $30\%$ SU(3) breaking on
penguin amplitudes (with flat distribution for the size of the SU(3) breaking) gives $\sqrt{\langle(\alpha^{\rm out}-\alpha^{\rm
in})^2\rangle}\sim 2^\circ$ \cite{Gronau:2004tm}.

In conclusion, time dependent measurements of $B(t)\to
\rho^\pm\pi^\mp$ allow for a determination of $\alpha$ already with
present data, if a mild assumption on the relative tree phase is used,
giving
\beq
\alpha=
(95 \pm 16)^\circ .
\eeq
With higher statistics a method that avoids even this mild assumption can be used. The theory error on extracted  value of $\alpha$ is very small due to small penguin
polution, i.e. $r_\pm\sim 0.2$.\\[-3mm]

I would like to thank M. Gronau for fruitful collaboration.

\end{document}